# ARTICLE  OPEN

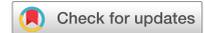

# Counterfactual ghost imaging

Jonte R. Hance [1✉] and John Rarity[1]

We give a protocol for ghost imaging in a way that is always counterfactual—while imaging an object, no light interacts with that object. This extends the idea of counterfactuality beyond communication, showing how this interesting phenomenon can be leveraged for metrology. Given, in the infinite limit, no photons ever go to the imaged object, it presents a method of imaging even the most light-sensitive of objects without damaging them. Even when not in the infinite limit, it still provides a many-fold improvement in visibility and signal-to-noise ratio over previous protocols, with over an order of magnitude reduction in absorbed intensity.

*npj Quantum Information* (2021)7:88 ; https://doi.org/10.1038/s41534-021-00411-4

## INTRODUCTION

Ghost Imaging exploits the the position-momentum entanglement between correlated photon pairs to derive image information. When one photon of the pair travels via the object and is focused into a bucket detector, an image can still be formed using coincident detection of the partner photon in a high-resolution pixel detector[1–4] (as shown in Fig. 1). While Pittman et al. originally conceived ghost imaging to illustrate the power of quantum correlations[5], it has since been shown thermal/classically correlated light can be used to replicate this effect classically[6–9]. However, doing this, rather than using a pair of entangled photons, removes some of the benefits of the original quantum protocol. Specifically, entangled pairs of photons created in spontaneous parametric downconversion (SPDC) are both correlated in terms of position (meaning they can image in the near-field), and anti-correlated in terms of momentum, meaning they are anti-correlated in position in the far-field, so can image there too, whereas classically, they can only be (anti-)correlated in one of the two conjugate variables, so can only image in one of these regimes. Therefore, despite the comparative ease of using thermal light, it makes more sense to image using entangled photon pairs. However, despite what may have been claimed[10], using short-wavelength light for the photons going to the high-resolution detector and longer wavelengths to the object does not allow an increase in imaging resolution above and beyond standard diffraction limits[11]. In any case, the ability to ghost image while reducing the energy going to the object under investigation, to reduce potential damage, could be massively beneficial[12,13].

Counterfactuality, an extension of interaction-free measurement is the idea of using quantum effects to either probe objects or send messages without any matter/energy passing between the two parties when information is transferred[14]. This is based on Elitzur and Vaidman's interaction-free bomb detector[15], where a bomb, set to trigger on detecting the presence of a single photon, is put in one of the arms of a Mach–Zehnder interferometer. Even if the bomb does not blow up, the field travelling (or blocked by the bomb) affects the interference pattern created at the output of the interferometer. The original protocol, using 50:50 beamsplitters, was inefficient, having a high probability of causing explosion. Since then, adaptions have been made that reduce the probability of the photon going via the bomb's path to nearly zero[16,17]. Extending the protocol to more mundane but realistic application scenarios,

Salih et al. created a communication protocol based on this idea, where Alice obtains a different result depending on whether or not Bob blocks his side of a channel, without any photon having gone via Bob[18,19] (which has even led to protocols which send quantum information counterfactually[20–23]). Given, in this protocol, the photon provably never travels via Bob when information is transmitted[24], it raises the question of whether this protocol could be adapted to probe an object counterfactually.

Zhang et al. combined ghost imaging with the Elitzur–Vaidman bomb detector[15] to create a form of ghost imaging where there is a chance that information is still received about the imaged object without any photons being absorbed by it[25]. However, the Elitzur–Vaidman bomb detector is not always necessarily counterfactual—there is a (reasonably high) chance the photon can go via the object being investigated[14]. By replacing the Elitzur–Vaidman object-detection system in Zhang's protocol with Salih et al.'s method for counterfactual communication, we create a protocol for ghost imaging that is always counterfactual—whenever information is received about the imaged object, no photons have interacted with that object.

Further, even in cases when no information travels, far fewer photons go to the object than in either the standard ghost imaging or in the interaction-free ghost imaging case—reducing the energy absorbed by the object, and so potentially damage done to that object by the imaging process.

## RESULTS

### Protocol

We first go through Salih et al.'s protocol for counterfactual communication, before showing how this can be adapted and integrated into Zhang et al.'s interaction-free ghost imaging protocol.

### Salih et al.'s Counterfactual Communication Protocol

Note, we define our polarisation Bloch sphere with poles $|H\rangle$ and $|V\rangle$, and rotation

$$\hat{\mathbf{R}}_y(\theta) = \begin{pmatrix} \cos(\frac{\theta}{2}) & -\sin(\frac{\theta}{2}) \\ \sin(\frac{\theta}{2}) & \cos(\frac{\theta}{2}) \end{pmatrix} = e^{-i\theta\hat{\sigma}_y/2} \qquad (1)$$

for dummy variable $\theta$ (in terms of Pauli-$y$ operator $\hat{\sigma}_y$).

---

[1]Quantum Engineering Technology Laboratory, Department of Electrical and Electronic Engineering, University of Bristol, Bristol, UK. ✉email: jonte.hance@bristol.ac.uk

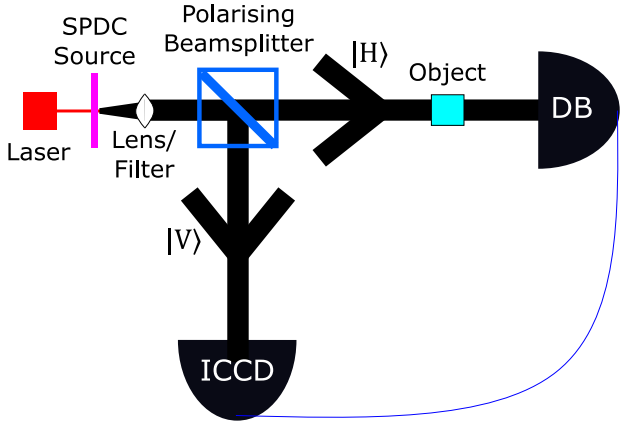

**Fig. 1 A standard set-up for ghost imaging.** Position-and-momentum-entangled photons are generated in pairs at the spontaneous parametric downconversion (SPDC) source from a laser beam, with conjugate polarisations (i.e. always in pairs of horizontal and vertical polarisation). A polarising beamsplitter sends the vertically polarised photon to an intensified charge-coupled device (ICCD) camera (which records in high resolution its arrival position) and sends the horizontally polarised photon via a sample to be imaged, to a bucket detector (DB). By recording coincidences between the detections at the ICCD and the bucket detector, the sample can be "ghost imaged".

A photon of state $a|H\rangle + b|V\rangle$ enters the outer interferometer through a half-wave plate (HWP), tuned to apply $\hat{\mathbf{R}}_y(\pi/M)$. It then enters a polarisation separator, subtly divert horizontally polarised light, and directly transmit vertically polarised light.

The vertical element then goes through the inner interferometer $N$ times. In each of these times it goes through a HWP tuned to apply $\hat{\mathbf{R}}_y(\pi/N)$, then through another polarisation splitter. The newly formed horizontal component passes across the channel, from Alice to Bob, who either blocks or does not block this channel. If he blocks, it is absorbed and lost—if not, it returns to Alice's side, recombines at another polarisation splitter with the vertical component, then is sent back into the inner interferometer by the switchable mirror. This happens $N$ times, before the switchable mirror is opened. The wave is then passed through one final polarisation splitter, which sends any horizontal components to loss detector DL, before being recombined at another polarisation splitter with the horizontal arm of the outer interferometer.

As each inner interferometer applies $\hat{\mathbf{R}}_y(\pi/N)$, if Bob does not block, the rotations sum to

$$\hat{\mathbf{U}}_{NB}^N = (e^{-i\pi\hat{\sigma}_y/N})^N = e^{-i\pi\hat{\sigma}_y} = \hat{\mathbf{R}}_y(\pi) \quad (2)$$

and so the state after the inner interferometer chain is

$$|V\rangle_I \to \hat{\mathbf{U}}_{NB}^N |V\rangle_I = |H\rangle_I \to \text{Loss} \quad (3)$$

The vertical component becomes horizontally polarised, and is lost to DL. Therefore, the only element of the wavefunction leaving the outer interferometer is that which travelled the outer arm.

Similarly, if Bob blocks for all inner interferometers, because of the quantum Zeno effect,

$$\hat{\mathbf{A}}_B^N = \left[ e^{-i\pi\hat{\sigma}_y/2N} \begin{pmatrix} 1 & 0 \\ 0 & 0 \end{pmatrix} \right]^N$$
$$= \begin{pmatrix} \cos\left(\frac{\pi}{2N}\right)^N & 0 \\ \cos\left(\frac{\pi}{2N}\right)^{N-1}\sin\left(\frac{\pi}{2N}\right) & 0 \end{pmatrix} \quad (4)$$

where $\hat{\mathbf{A}}_B^N$ is non-unitary. Therefore, the state after the outer interferometer is

$$|V\rangle \to \hat{\mathbf{A}}_B^N |V\rangle_I$$
$$= \cos\left(\frac{\pi}{2N}\right)^N |V\rangle_I + \cos\left(\frac{\pi}{2N}\right)^{N-1}\sin\left(\frac{\pi}{2N}\right)|H\rangle_I \quad (5)$$
$$\to \cos\left(\frac{\pi}{2N}\right)^N |V\rangle + \text{Loss}$$

so some vertically polarised component exits the outer interferometer.

This means the outer cycle applies

$$\begin{pmatrix} 1 & 0 \\ 0 & 0 \end{pmatrix} e^{-i\pi\hat{\sigma}_y/2M} \quad (6)$$

if Bob does not block, or

$$\begin{pmatrix} 1 & 0 \\ 0 & \cos\left(\frac{\pi}{2N}\right)^N \end{pmatrix} e^{-i\pi\hat{\sigma}_y/2M} \quad (7)$$

if he does. They repeat this $M$ times, starting with a horizontally polarised photon, and using a final polarisation splitter to split it into horizontal and vertical components.

Because Alice applies $\hat{\mathbf{R}}_y(\pi/M)$ at the start of each outer interferometer, if Bob does not block, the state of the photon after $M$ outer cycles is

$$\cos\left(\frac{\pi}{2M}\right)^M |H\rangle \quad (8)$$

meaning, if it is not lost, it remains horizontally polarised (and goes into $D_0$). However, if he blocks, the state of the photon after $M$ outer cycles is (as $N \to \infty$) is $|V\rangle$, and it becomes vertically polarised (and goes into $D_1$).

## Adapting to ghost imaging

We now describe how to adapt the above protocol to use it for counterfactual ghost imaging (as shown in Fig. 2). A pair of conjugately polarised photons, entangled in position and momentum, is split at a polarising beamsplitter. The vertically polarised photon is sent through an image-preserving optical delay line to an Intensified Charge-Coupled Device (ICCD) camera (both of which have previously been used to allow multi-mode ghost imaging, rather than single-mode raster-scanning[3,10,26]). The horizontally polarised photon is sent through one run of Salih et al.'s protocol, where the object to be imaged is put in place of Bob's blocker. If the object does not block the path, the photon, in that spatial mode, remains horizontally polarised, and so goes to $D0$, leading to a coincidence measurement between $D0$ and that pixel of the ICCD camera; however, if the object does block the path, the photon becomes V-polarised and so goes to $D1$, causing a coincidence measurement between $D1$ and that pixel of the ICCD camera.

Note, because arrival in $D0$ ($D1$) is correlated far more closely in Salih et al.'s protocol with the object not blocking (blocking) the path (see Fig. 3b) than in the Elitzur–Vaidman bomb detector (and so Zhang et al.'s protocol[25]), we only need to resolve coincidences between $D0$ and the ICCD (unlike Zhang et al., who need to form two images - one based on $D0$-ICCD coincidences, and the other on $D1$-ICCD coincidences—and subtract one from the other, due to the interference patterns produced by the Elitzur–Vaidman bomb detector[15]). However, by resolving both $D0$-ICCD and $D1$-ICCD coincidences, and subtracting one from the other, we can image with high accuracy even for low $N$ - so we do this.

## Use of other counterfactual protocols

A similar protocol could be constructed by replacing Salih et al.'s protocol in the imaging set-up with either Aharonov and Vaidman's modified protocol[27], or Vaidman's later adaptation[28]—however, such a protocol would not be counterfactual by the consistent histories criterion (as shown by Salih[21,24]).

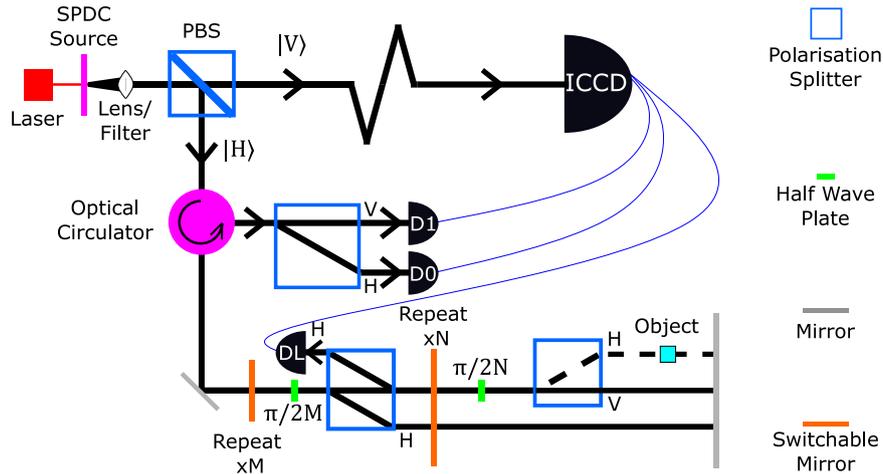

**Fig. 2 Our counterfactual ghost imaging protocol.** This is based on the combination of standard ghost imaging (Fig. 1) and a common-path interferometer version of Salih et al.'s counterfactual communication protocol[18]. We create a pair of position-and-momentum-entangled photons, one horizontally polarised and one vertically polarised, by passing a pulsed pump laser through a spontaneous parametric downconversion (SPDC) crystal, before collimating the beam, and filtering out the pump. The photon pair is split at a polarising beamsplitter (PBS), with the $V$-polarised photon going through a long optical delay to an Intensified Charge-Coupled Device (ICCD) camera, and the $H$-polarised photon going through a run of Salih et al.'s protocol, adapted so the object to be investigated is put in place of Bob's blocker. The switchable mirrors allow the photon to cycle the correct number of times: the first for $M$ outer cycles; and the second for $N$ inner cycles per outer cycle. The polarisation separators subtly divert horizontally polarised light, and directly transmit vertically polarised light. The half-wave plates are tuned to implement a $\hat{\mathbf{R}}_y(\theta)$ polarisation-mode rotation with $\theta$ of $\pi/2M$ and $\pi/2N$, respectively. The detector DL acts as our loss channels (which we postselect against). After M outer cycles, the switchable mirror sends the photon to the optical circulator, which sends it to the PBS. The path not being blocked by the object leads the photon to remain $H$-polarised, and so go to the ICCD, leading to a coincidence measurement between that and the ICCD camera; however, the path being blocked leads to the photon becoming $V$-polarised and so going to $D1$, so coincidence measurement between that and the ICCD. The use of multi-mode interferometers and (position-momentum) correlations between the entangled photons enables multi-mode ghost imaging in this counterfactual set-up. Note, the polarisation separators ensure a common path length for both $H$- and $V$-polarised components, while generating beam separations of the several millimetres. An optimisation we mention in the discussion has photons going to DL can trigger a coincidence measurement with the ICCD, treated as if it was a detection at $D0$, which does not affect the chance of photons interacting with the object (photons only go to DL if the object does not block the channel), and allows us to lower the number of outer cycles to the minimum required (2) with no increase in loss.

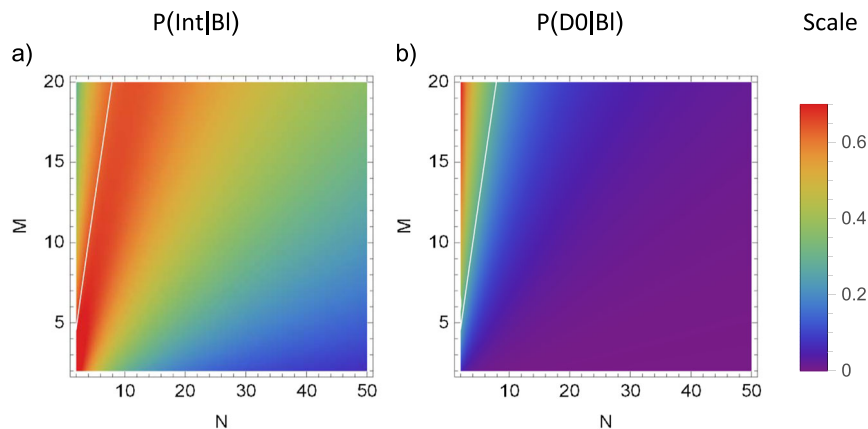

**Fig. 3 Loss probabilities when the object blocks the path.** Probability, when the object blocks the channel, of the photon: interacting with the object being imaged ($P_{\text{int}}$) (**a**); or erroneously ending up in $D0$ (**b**). We plot these for given numbers of outer ($M$) and inner ($N$) interferometer cycles. Note, the photon only goes via the object erroneously - in any case when $D1$ clicks, the detection of the object will have been fully counterfactual. Further, both the interaction and erroneous-$D0$ probabilities go to 0 as $N$ goes to $\infty$.

### Generation of position-entangled photon pairs

For the original ghost imaging protocol, photon pairs are generated by SPDC[5]. This makes use of second-order nonlinearities in an optical medium to generate conjugately polarised photon pairs entangled in position and with frequencies that sum to the frequency of an input pump laser. By using low-pass filters, the photons can be split off from the pump laser, and then split from one another using a polarising beamsplitter, sending one to the high-resolution detector, and the other to the object and bucket detector. We propose using the same source for our protocol.

Also note, for the $V$-polarised photon going to the ICCD, rather than using an optical delay, the ICCD can detect the photon earlier, but record the time of arrival as well as the position, which can be used with post-processing to determine effective coincidences with the bucket detectors, avoiding issues with long optical delays.

## DISCUSSION

Counterfactuality has been a controversial subject, with many debating whether or not given protocols are counterfactual—most notably from the weak trace[29] and consistent histories[30] approaches to the path of a quantum particle. However, unlike other protocols[27,28,31–33] Salih et al. have recently shown their protocol for counterfactual communication is counterfactual by both these criteria[24]—whenever Alice detects the photon at either $D0$ or $D1$, she can be sure it has never been at Bob. However, when the number of cycles is not infinite, there is a chance the photon could end at Bob, rather than Alice, in which case the protocol is aborted and restarted. Given a use of ghost imaging is to image photosensitive samples[12,13] (which could be easily damaged by high-energy photons), we want to reduce the chance of any photons going to/via the object as much as possible. We plot this probability in Fig. 3a. Note, as $N$ goes to infinity, this probability goes to zero.

Interestingly, Aharonov and Rohrlich have recently shown modular angular momentum $L_z \mathrm{mod} 2\hbar$ of $\hbar$ is conveyed by counterfactual communication from Alice's photon into Bob's blocker whenever Bob blocks[34]—however, there is no energy associated with this, meaning there is no chance of this damaging photosensitive samples; therefore, we can ignore this in our analysis.

In Salih et al.'s 2013 protocol, there is a probability of erroneous $D0$ clicks, as they take $\cos{(\pi/2N)}^N \to 1$ for large $N$. This probability, for $M = 2$, is

$$P(D0|\text{Block})_{M=2} = \left( \cos\left(\frac{\pi}{2N}\right)^N - 1 \right)^2 / 4 \qquad (9)$$

We plot the probability for given values of $M$ and $N$ in Fig. 3b. By increasing the rotation slightly at the start of each outer cycle, this error could be avoided—future work will consider the exact rotation needed, and the specific benefits of this optimisation.

The signal-to-noise ratio (SNR)[12,35,36], a useful measure of the efficacy of an imaging system, is given by

$$\text{SNR} = \frac{|\Delta \bar{I}|}{\sigma(|\Delta \bar{I}|)} \qquad (10)$$

where $\Delta \bar{I}$ is the difference in average intensity values observed by a detector between inside and outside the object, and $\sigma(|\Delta \bar{I}|)$ is the standard deviation in this difference.

For standard ghost imaging, when an average of $\bar{N}$ photons (those generated in a given time interval by a SPDC source) interrogate an object, none of the $\bar{N}$ photons will reach the detector, giving a change in photon detection number at the detector of $\Delta N_{GI} = -\bar{N}$. Given SPDC has thermal statistics for the rate of emission (which look Poissonian averaged over many temporal modes), the SNR is

$$\text{SNR}_{GI} = \bar{N}/\sqrt{\bar{N}} = \sqrt{\bar{N}} \qquad (11)$$

For counterfactual ghost imaging, we define $\Delta N_{D0}$ ($\Delta N_{D1}$) as the difference in photon numbers received at $D0$ ($D1$) between the object blocking and not blocking the channel (which in each case is $\bar{N}$ times the difference in probability of a photon reaching that detector in each of those two cases). Note, $\Delta N_{D0}$ and $\Delta N_{D1}$ will have opposite signs. Therefore,

$$\begin{aligned} \text{SNR}_{CGI} &= \frac{|\Delta N_{D0} - \Delta N_{D1}|}{\sigma(|\Delta N_{D0} - \Delta N_{D1}|)} \\ &= f(M,N)\sqrt{\bar{N}} = f(M,N)\text{SNR}_{GI} \end{aligned} \qquad (12)$$

which we plot in Fig. 4a (as a multiple of $\text{SNR}_{GI}$, the SNR of standard ghost imaging). For values of $M$ and $N$ where $\text{SNR}_{CGI} = 1$, the protocol is just as good at imaging as standard ghost imaging. In these cases, Fig. 3b shows that in our protocol the probability of a photon interacting with the object is much less than the 73.5% limit from previous protocols[25].

However, rather than looking at the SNR for the photons generated by a SPDC source in a given period of time, a more apt comparison would be the SNR for which the same number of photons are absorbed by the object as in standard ghost imaging. Given the average number of photons interacting with the object is $P_{\text{Int}}$ times $\bar{N}$, we get

$$\text{SNR}_{\text{int}} = f(M,N)\sqrt{\bar{N}/P_{\text{Int}}} = \frac{f(M,N)}{\sqrt{P_{\text{Int}}}}\text{SNR}_{GI} \qquad (13)$$

which we plot in Fig. 4b (again in terms of $\text{SNR}_{GI}$).

Even for low numbers of outer cycles ($M$ and $N$), our protocol gives a vast improvement over the SNR of standard ghost imaging—for instance, two outer cycles of 13 inner cycles gives double the equal-photon-absorption SNR of standard ghost imaging. This is much higher than the 118% improvement in equal-photon-absorption SNR that Zhang et al.'s protocol gives[25]. Note, as $N \to \infty$, the probability of a photon interacting with the object goes to 0, meaning $\text{SNR}_{\text{int}}$ becomes infinitely larger than the SNR available with standard ghost imaging (if we are willing to wait that long).

Another key measure of an imaging protocol's efficacy is its visibility, $V$, defined as

$$V = \frac{|\bar{N}_{\text{In}} - \bar{N}_{\text{Out}}|}{\bar{N}_{\text{In}} + \bar{N}_{\text{Out}}} \qquad (14)$$

Visibility gives how responsive to a difference in presence/absence of an object is, defined on a scale from 0 to 1. For our protocol, visibility is given as

$$V_{CGI} = \frac{|\Delta N_{D0} - \Delta N_{D1}|}{2\bar{N}} \qquad (15)$$

(the changes in intensity for $D0$ and $D1$ over the maximal possible changes in their intensities, remembering their opposite signs). This gives reasonable values (i.e., between 0 and 1) for our protocol, as given in Fig. 4c—for instance, 5 outer cycles of 12 inner cycles gives a visibility of 0.569. This shows higher visibility than the maximal we calculate for Zhang's protocol (0.5625)[25].

When the object does not block the photon's path in the adapted Salih et al. device, for finite numbers of outer interferometers ($M$), there is a chance the photon will cross through the unblocked gap, in which case it goes to the loss detector DL rather than the coincidence-linked detector $D0$. This occurs with probability

$$P(\text{DL}|\text{NB}) = 1 - \cos\left(\frac{\pi}{2M}\right)^{2M} \qquad (16)$$

which we plot in Fig. 5. This goes to 0 as $M$ goes to infinity.

Note, if we weaken our definition of counterfactuality to be that the photon never goes via the object's path when the object is there (rather than the photon never goes via that path at all), we could link detector DL as well as $D0$ to the ICCD, treating coincidences between DL and the ICCD as if they were between $D0$ and the ICCD. This would let us use the minimum number of outer cycles the protocol works for, two, rather than requiring higher values of $M$ to avoid us erroneously ignoring ICCD detections. We do a version of this in Fig. 2, having photons which would go to DL treated as if going to $D0$.

This led us to plot altered values for SNRs and visibility, taking into account DL now going to $D0$. These are plotted in Fig. 4 (d, e, f), and give even better values for all three measures.

Traditional ghost imaging can only distinguish between whether or not a photon could have passed through a given region (i.e., whether, at that point, a mask would transmit that photon, or whether it would absorb/reflect it). This makes this style of ghost imaging bad for imaging transparent/translucent objects, which is unfortunate, given the many applications for the detection of low-contrast objects. However, Abouraddy et al., and later Gong et al., proposed[37,38] (and Zhang et al. experimentally

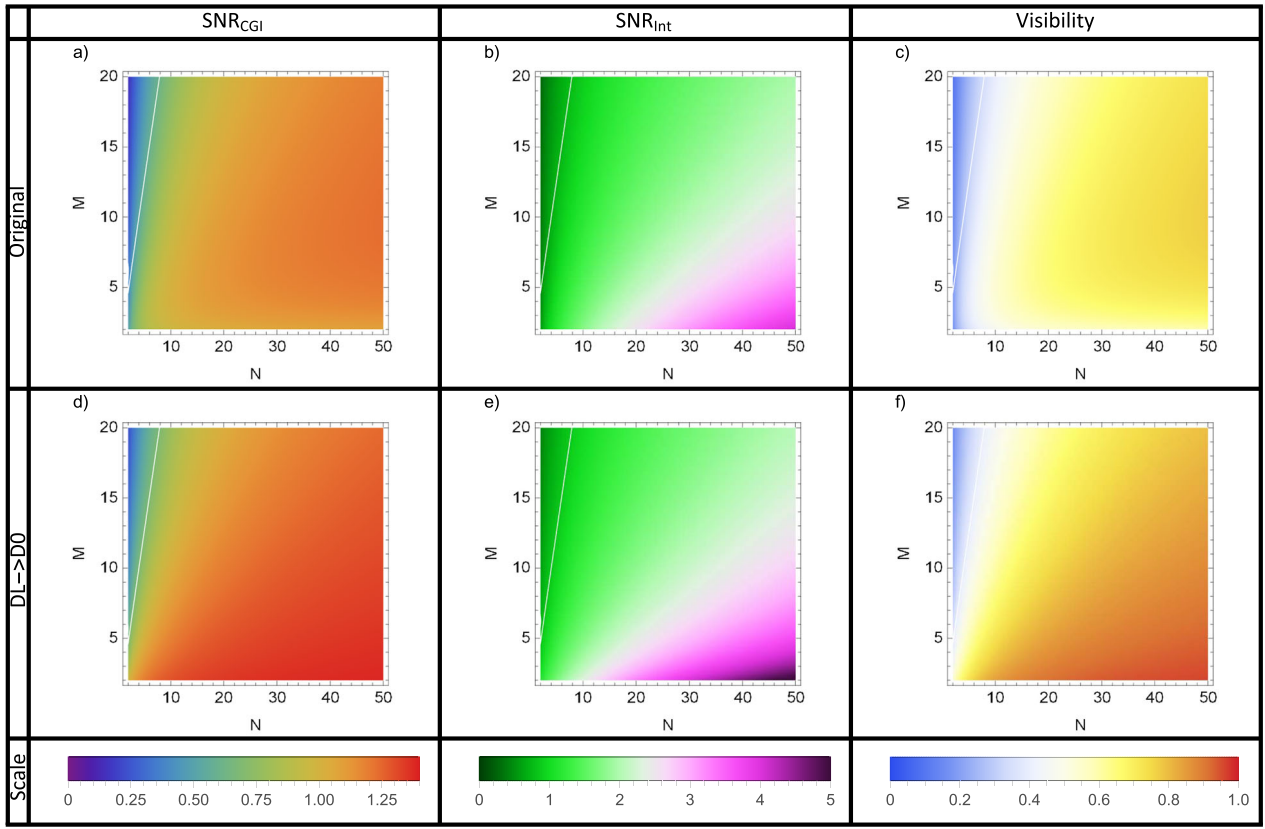

**Fig. 4 Plots of the signal-to-noise ratio (SNR).** This is for: equal numbers of photon pairs generated in a given time by the SPDC source (**a**); and equal numbers of photons absorbed by the object (**b**)—and the visibility V of the protocol (**c**)—for our original protocol; and these values for when photons that would go to DL also count for coincidence measurements as if they went to D0 (**d**, **e** and **f**). These are as functions of the number of outer (M) and inner (N) interferometer cycles.

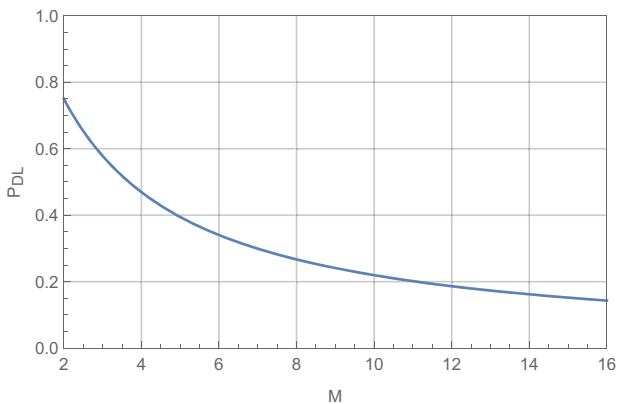

**Fig. 5 Probability the photon goes to DL rather than D0.** This is the probability that, when the object does not block the photon' path, the photon travels via that path and so goes to DL rather than D0. This is as a function of number of outer cycles (M).

demonstrated[39]) schemes, which make the use of phase-sensitivity to allow materials to be detected which, while translucent, create a change to the phase/polarisation of transmitted light. Zhang et al.'s interaction-free ghost imaging also demonstrated this sensitivity[25]. Further, given it relies on interference to direct the photon to one of two bucket detectors, our scheme is also sensitive to changes in phase induced by transparent objects - presenting yet another benefit of our protocol over standard ghost imaging.

Alongside standard ghost imaging, which makes use of entangled photons, alternative forms have been created which instead just use classical correlations. Given this is classically simulable, a version has been demonstrated which is referred to as computational ghost imaging—where rather than sending a correlated photon to a high-resolution/scanning detector, a spatial light modulator (SLM) applies an effective reference pattern to the photon probing the object. Given counterfactual ghost imaging allows us to image an object counterfactually while preserving quantum correlations between the signal and idler photons, it can clearly preserve the classical correlations necessary for computational ghost imaging. In such a set-up, we replace the SPDC, beamsplitter and ICCD camera with a single photon source, a SLM, a pseudo-random illumination pattern, and computational analysis. While it remains to be seen the effect the loss of quantum correlations would have on fidelity, loss, SNR and visibility, this shows the flexibility of counterfactual alterations to ghost imaging.

The analysis presented above assumes ideal components. Sadly, no component is ideal. In our protocol, the four key components which could through loss affect the protocol are the HWPs, the polarising beamsplitters, the switchable mirrors, and the detectors. In this appendix, we model the protocol with these experimentally realistic values, to show that, even with these limitations considered, the protocol still provides a significant advantage over both standard ghost imaging, and classical metrology.

At designed-for wavelengths, HWPs can achieve loss (through reflection) of $\mathcal{O}(0.1\%)$; polarising beamsplitters can achieve loss (through absorption) of less than 1%, and a typical heralding efficiency for a SPDC/SPAD set-up like ours is 18%. Practically, switchable mirrors pose an issue (given typical switching times for these are of $\mathcal{O}(10^{-6}s)$, while the switching we need has to be of $\mathcal{O}(10^{-9}s)$). An issue with this is lossiness (adding loss in their demonstration of 15/16 per outer cycle). Despite this, as we show in Fig. 6 (which shows the same quantities as in Fig. 4 albeit adjusted to take into account these losses), the SNR per photon absorbed is still far higher than for standard ghost imaging.

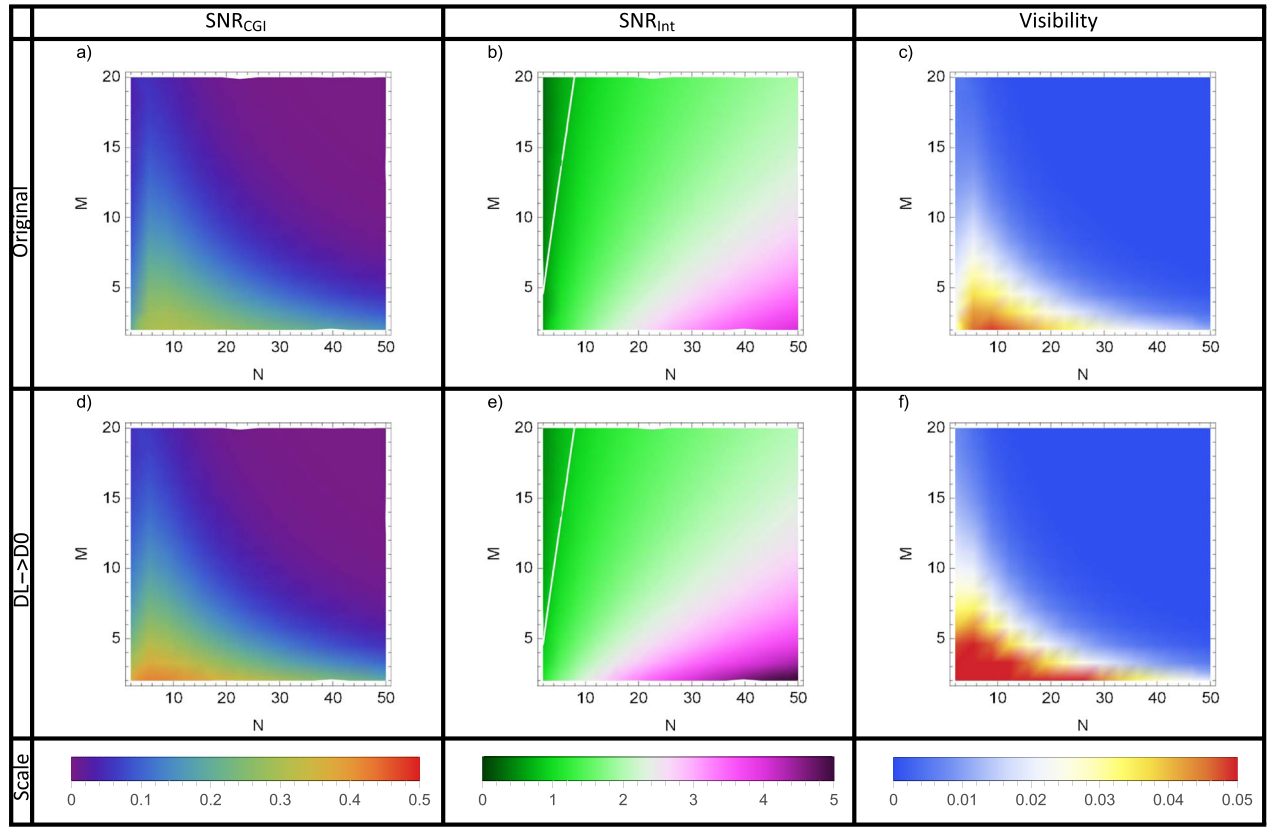

**Fig. 6 Plots (assuming realistic component loss) of the signal-to-noise ratio (SNR).** These are for: equal numbers of photon pairs generated in a given time by the SPDC source (**a**); and equal numbers of photons absorbed by the object (**b**)—and the visibility $V$ of the protocol (**c**)—for our original protocol; and these values for when photons that would go to DL also count for coincidence measurements as if they went to $D0$ (**d**, **e** and **f**). These are as functions of the number of outer ($M$) and inner ($N$) interferometer cycles.

We have given a protocol for ghost imaging in a way that is always counterfactual—while imaging the object, no light interacts with that object. This extends the idea of counterfactuality beyond communication, showing how this interesting phenomenon can be used for metrology. Given, in the infinite limit, no photons ever go to the imaged object, it presents a method of imaging even the most light-sensitive of objects without damaging them. A future direction will be looking at separately leveraging the chained quantum Zeno effect to see if this performance can be improved further.

### DATA AVAILABILITY
Data sharing not applicable to this article as no datasets were generated or analysed during the current study.

### CODE AVAILABILITY
The code generated to analyse the protocol is available from the corresponding author upon reasonable request.

## ACKNOWLEDGEMENTS
We thank Hatim Salih, Alex McMillan and Will McCutcheon for useful discussions. This work was supported by the University of York's EPSRC DTP grant EP/R513386/1, and the Quantum Communications Hub funded by the EPSRC grant EP/M013472/1.

## AUTHOR CONTRIBUTIONS
J.R.H.—conceptualisation, methodology, software, formal analysis, writing (original draft), visualisation; J.R.— conceptualisation, supervision, writing (review and editing), funding acquisition.

## COMPETING INTERESTS
The authors declare no competing interests.

## ADDITIONAL INFORMATION
**Correspondence** and requests for materials should be addressed to J.R.H.

**Reprints and permission information** is available at http://www.nature.com/reprints

**Publisher's note** Springer Nature remains neutral with regard to jurisdictional claims in published maps and institutional affiliations.